\documentclass[
aps,%
11pt,%
final,%
notitlepage,%
oneside,%
twocolumn,%
nobibnotes,%
nofootinbib,%
superscriptaddress,%
noshowpacs,%
centertags]%
{revtex4}

\usepackage{multirow}
\renewcommand{\baselinestretch}{1.0}

\begin{document}

\selectlanguage{english}


\title{Ultra-Flat Galaxies Selected from RFGC Catalog.
I. The Sample Properties}

\author{\firstname{V.~E.}~\surname{Karachentseva}}
\affiliation{Main Astronomical Observatory of National Academy of
Sciences of Ukraine, Kiev, 03680 Ukraine}

\author{\firstname{Yu.~N.}~\surname{Kudrya}}
\affiliation{Astronomical Observatory of Taras Shevchenko National
University of Kiev, Kiev,  04053 Ukraine}

\author{\firstname{I.~D.}~\surname{Karachentsev}}
\email{ikar@sao.ru} \affiliation{\saoname}

\author{\firstname{D.~I.}~\surname{Makarov}}
\email{dim@sao.ru} \affiliation {\saoname}

\author{\firstname{O.~V.}~\surname{Melnyk}}
\affiliation{Taras Shevchenko National University of Kiev, Kiev,  01033 Ukraine}

\received{October 27, 2015}  \revised{December 22, 2015}

\onecolumngrid
{\scriptsize
ISSN 1990-3413, Astrophysical Bulletin, 2016, Vol. 71, No. 1, pp.1-13 @ Pleiades Publishing, Ltd., 2016\\
Original Russian Text @ V.E.Karachentseva, Yu.N.Kudrya, I.D.Karachentsev, D.I.Makarov, O.V.Melnyk, 2016,\\
 published in Astrofizicheskii Byulleten, 2016, Vol.71, No.1, pp.1-13.}

\begin{abstract}
We used the Revised Flat Galaxy Catalog (RFGC) to create a sample of
ultra-flat galaxies (UFG) covering the whole northern and southern
sky apart from the Milky Way zone. It contains 817 spiral galaxies
seen edge-on, selected into the UFG sample according to their
apparent axial ratios $(a/b)_B\geq10.0$ and $(a/b)_R\geq8.53$ in the
blue and red bands, respectively. Within this basic sample we fixed
an exemplary sample of 441 UFG galaxies having the radial velocities
of $V_{LG} < 10000$ km s$^{-1}$, Galactic latitude of $\mid
b\mid>10^{\circ}$ and the blue angular diameter of $a_B > 1\farcm0$.
According to the Schmidt test
the exemplary sample of 441 galaxies is characterized by about (80--90)\%
completeness, what is quite enough to study different properties of the ultra-flat
galaxies. We found that more than 3/4 of UFGs have the morphological types within the
narrow range of $T= 7\pm1$, i.e. the thinnest stellar disks occur among the Scd, Sd,
and Sdm types. The average surface brightness of UFG galaxies tends to diminish towards
the flattest bulge-less galaxies. Regularly shaped disks without signs of
asymmetry make up about 2/3 both among all the RFGC galaxies, and the UFG sample objects.
About 60\% of ultra-flat galaxies can be referred to dynamically isolated objects,
while 30\% of them probably belong  to the scattered associations (filaments,
walls), and only about 10\% of them are dynamically dominating  galaxies with respect to
their neighbours.
\end{abstract}

\maketitle

\section{INTRODUCTION}
 In his study ``The Classification of Spiral
Galaxies''~\cite{hub1927:Karachentseva_n} Edwin Hubble, answering the
criticisms of John Reynolds has re-designated the structural features
that divide the spiral galaxies into the ``early''  (Sa),
``intermediate'' (Sb) and ``late'' (Sc) types. He called the relative
size of the unresolved galactic nucleus (sign 1) the first and basic
criterion, i.e. in the contemporary terminology---the size of the
bulge with respect to the disk. By the time, only 290 images were
used to classify the spirals, and hence, Hubble did not consider  the
axial ratio to be an essential criterium for the classification, as
urged by Reynolds.  Nevertheless, sign~(1), along with the openness
of the spiral structure~(2) and the degree of concentration of matter
in the arms~(3) formed a solid basis for Hubble's statistical
conclusions about the relationship of different characteristics of
galaxies. In the same paper~\cite{hub1927:Karachentseva_n} Hubble did
not agree  with Reynolds's assumption that the spirals, seen edge-on,
should be classified in a separate class, according to the ratio of
their axes and the absorption pattern   in them.


A further elaboration and development of the Hubble sequence of galaxies is
due mainly to the widely known studies of  de Vaucouleurs and  Sandage
(see, e.g., the survey by Buta~\cite{but1989:Karachentseva_n}).
The class of the late Hubble spirals Sc got a natural extension for the
``bulge-less''  Scd, Sd, and Sm types for all the angles of inclination to the line of sight $i$.
 At that, as shown by Karachentsev~\cite{kar1989:Karachentseva_n},
the galaxies, seen  nearly edge-on  ($i > 85\degr$) are much easier
to classify by the (reverse)  ``bulge/disk'' ratio and isolate among
them   very thin purely disk spirals. As it was stressed by Kormendy
and  Kennicutt~\cite{kor2004:Karachentseva_n}, ultra-thin spiral
galaxies are of particular interest in light of their origin and
survival in the environments of   different density.


Flat galaxies were long known as spiral galaxies of late
morphological types, seen edge-on,   with a small or a missing
nucleus~\cite{vor1974:Karachentseva_n,hei1972:Karachentseva_n}. To
date,
 a large number of observations of flat galaxies were conducted in the
optical and radio ranges (see the survey of
Kautsch~\cite{kau2009:Karachentseva_n}). However, a systematic
catalogization of flat galaxies became possible only at the advent of
homogeneous sky surveys, implementing   certain selection conditions.


The first catalog of flat edge-on galaxies, the Flat Galaxy Catalog,
FGC(E) and its updated RFGC version are published
in~\cite{kar1993:Karachentseva_n,kar1999:Karachentseva_n}. The RFGC
catalog~\cite{kar1999:Karachentseva_n} covers the entire sky and
contains 4236 galaxies, visually selected on the blue (further, $B$)
and red (further, $R$) prints of the first Palomar sky survey POSS-I
and the  ESO/SERC survey, with the ``blue'', in the POSS-I system,
major axis  $(a/b)_B \geq 7$  at the angular diameter (major axis) of
$a_B\geq0\farcm6$.  Further, for brevity, we shall use the
expressions ``blue (red) diameter'' and ``blue (red) axial ratio'',
referring to the diameter and the galaxy image axial ratio   on the
blue (red) print. The criterion of  the axial ratio was later used as
one of the main factors  creating the catalogs of flat spiral
galaxies: in the near infrared 2MASS
range~\cite{jar2000:Karachentseva_n}, in
2MFGC~\cite{mit2004:Karachentseva_n} and in different releases of the
Sloan survey: SDSS DR1~\cite{kau2006:Karachentseva_n} and SDSS
DR7~\cite{biz2014:Karachentseva_n}.


In his review paper~\cite{kau2009:Karachentseva_n} Kautsch describes
various models of formation and evolution of thin disks, the
properties of such objects based on the data from the
catalogs~\cite{kar1993:Karachentseva_n,kar1999:Karachentseva_n,kau2006:Karachentseva_n},
as well as the results of observations of individual objects. The
details can be found in the  extensive list of references therein.
Note that the choice of the  apparent axial ratio criterion for
ultra-thin galaxies by different authors is rather arbitrary (see the
survey of observational data in~\cite{kau2009:Karachentseva_n}). For
example, Goad and Roberts~\cite{goa1981:Karachentseva_n} give the
results of spectroscopic  observations for the galaxies with $a/b$ in
the range of 9--20 (the optical range).


The well-known ``classical'' super-thin isolated
galaxy   UGC\,7321 = FGC\,1403 = RFGC\,2246 = 2MFGC\,9681
\linebreak (KIG\,524~\cite{kara1973:Karachentseva_n}= 2MIG\,1699~\cite{kara2010:Karachentseva_n})
 has an axial ratio of   $(a/b)_B = 16, (a/b)_R = 13$ in the RFGC.  Meanwhile, the
  H$\alpha$ line observations  for the  subsystem of the emission
 H\,II regions of this galaxy give the axial ratio of $(a/b)_{{\rm
H}\alpha} = 38$ (see Fig.~1).


These numbers show that the axial ratio  of a galaxy  depends on the
color (age) of its stellar population. The flattest subsystem is
formed by the youngest stars with the age of about 10~Myr,
concentrated in the H\,II regions.

\begin{figure} \setcaptionmargin{5mm} \onelinecaptionstrue
\captionstyle{normal}
\includegraphics[width=1.1\columnwidth]{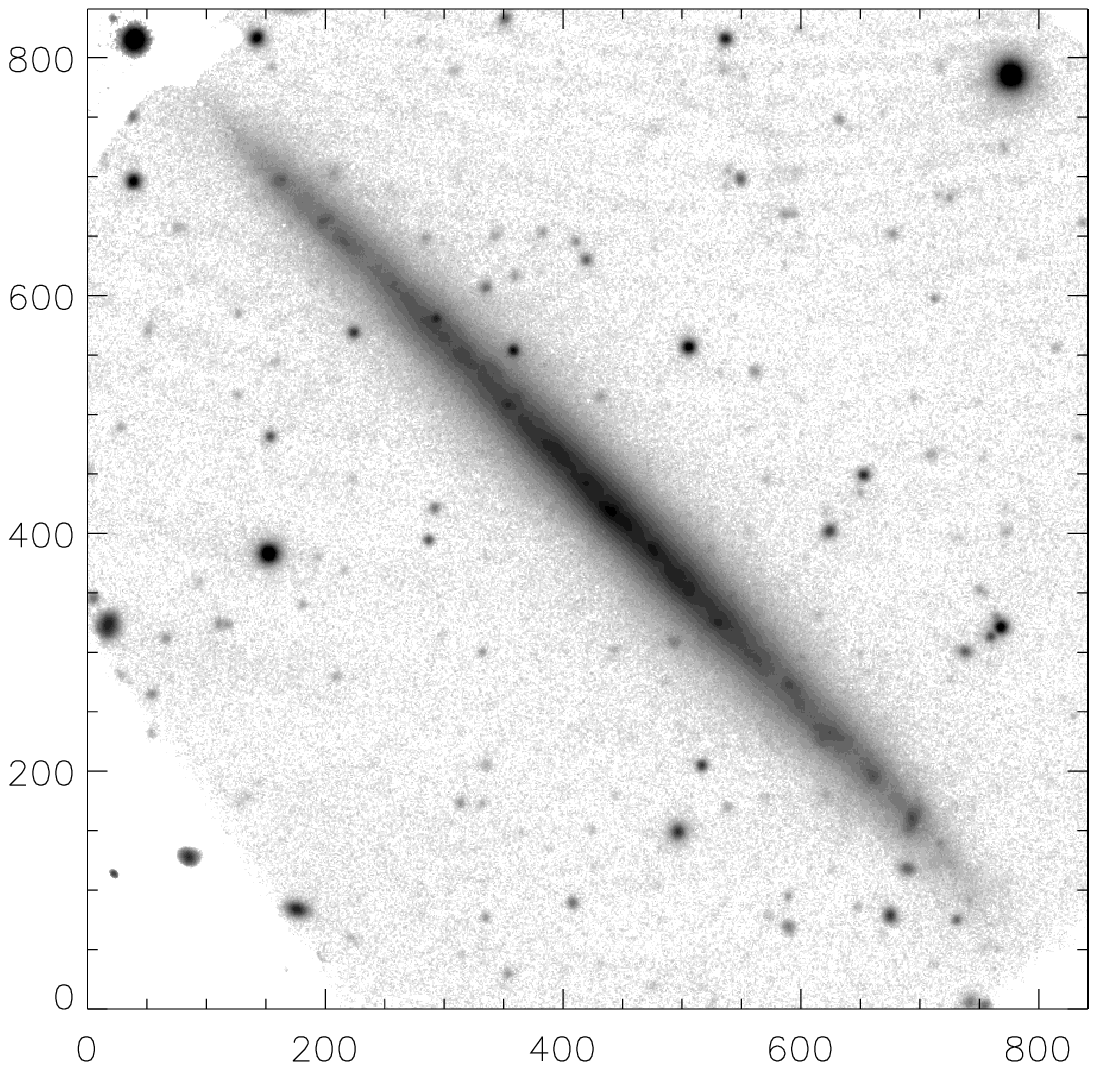}
\includegraphics[width=1.1\columnwidth]{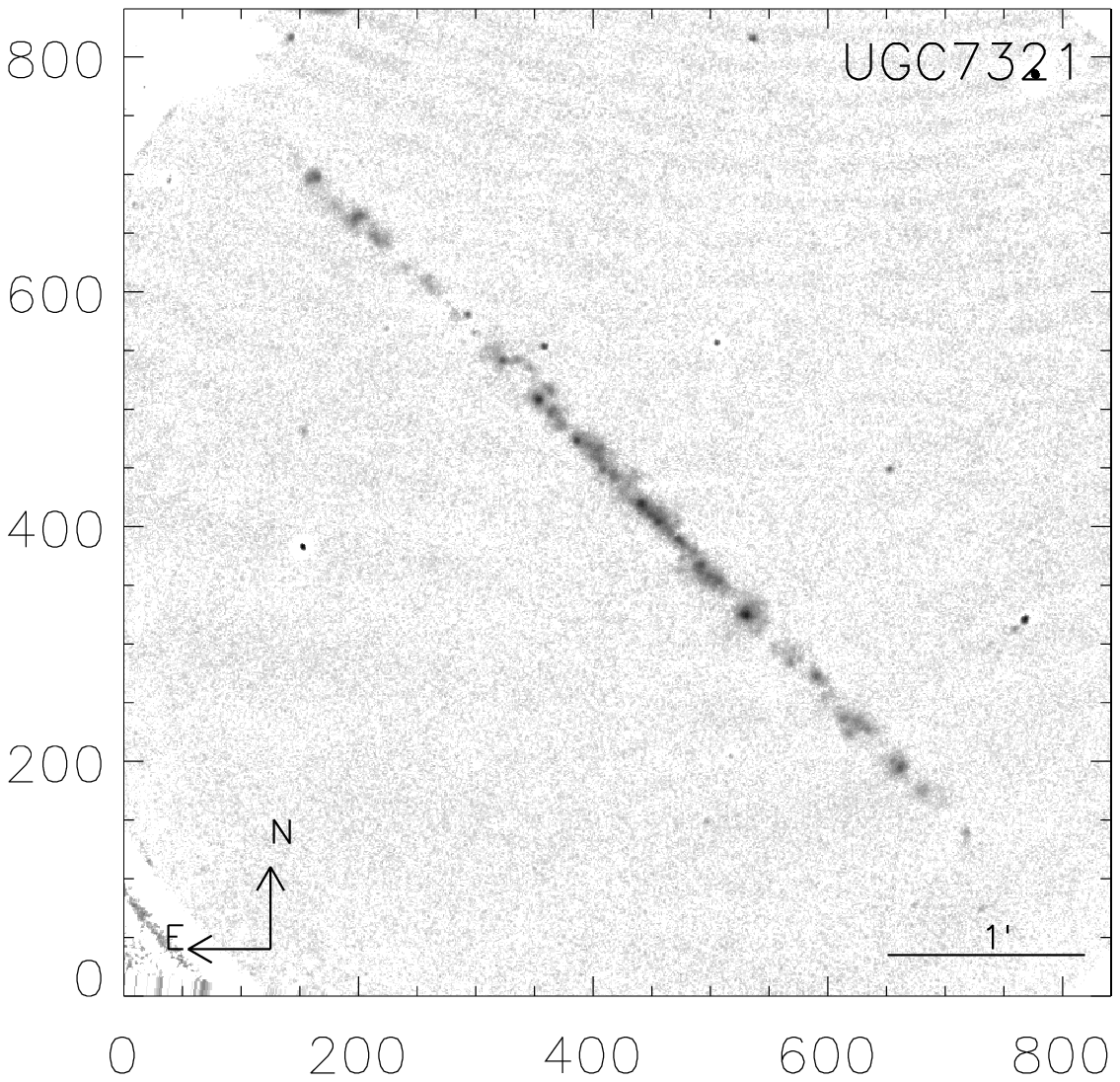}
\caption{A pair of images of the UGC\,7321 galaxy, obtained at the
6-meter  BTA telescope with the  SCORPIO focal
reducer~\cite{kar2015:Karachentseva_n}. Top: an image  in the
continuum with  the SED607+SED707 filters, $a/b=14$. Bottom: an image
in the line of H$\alpha$ with the subtraction of the continuum,
$a/b=38$. The scale and orientation are shown in the  corners of the
bottom image.}
\end{figure}

For all the galaxies in the  FGC(E) catalog, the apparent axial ratio  in the blue
 range does not exceed   $(a/b)_B = 22.4$. It corresponds to the maximum value of true
(spatial) axial ratio of 25.8~\cite{kud1994:Karachentseva_n}. The latter value is of
vital importance  for the models of formation and stability of thin
stellar disks.


The aim of this study is to create an exemplary  sample of ultra-thin
galaxies from the RFGC catalog and compare  the presented properties
of such objects, located in different environments. Section~2 briefly
enumerates selection effects, affecting the image of a spiral galaxy
seen edge-on, and describes the selection procedure of ultra-thin
galaxies from the  RFGC catalog. Section~3 shows the characteristics
of ultra-thin galaxies compared with all the RFCG galaxies. In
Section~4, we  consider different ways of identifying the environment
and compare the catalog properties of ultra-thin galaxies, located in
different environments. Brief conclusions are given in Section~5.

\section{A SAMPLE OF SUPER-THIN EDGE-ON SPIRAL GALAXIES}

\begin{figure}
\setcaptionmargin{5mm} \onelinecaptionstrue \captionstyle{normal}
\includegraphics[width=\columnwidth]{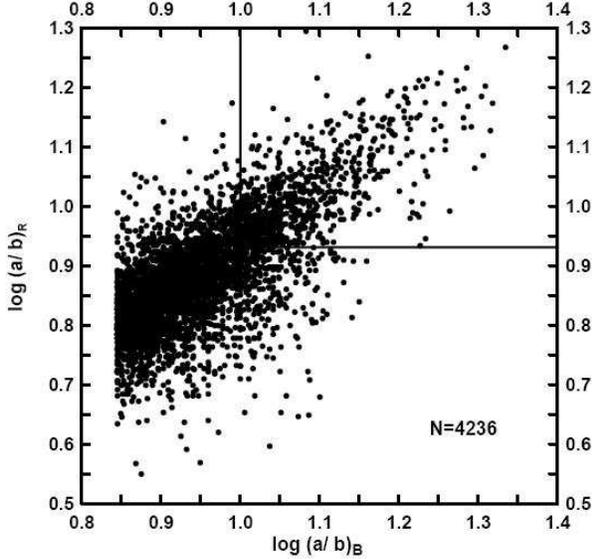}
\caption{The distribution of RFGC galaxies by the red  $(R)$  and
blue ($B$) axial ratios. The upper right corner contains the UFG
galaxies, limited by the lines  $(a/b)_B = 10$ and $(a/b)_R = 8.53$.}
\end{figure}

\begin{figure*}
\setcaptionmargin{5mm} \onelinecaptionstrue \captionstyle{normal}
\includegraphics[width=0.475\textwidth]{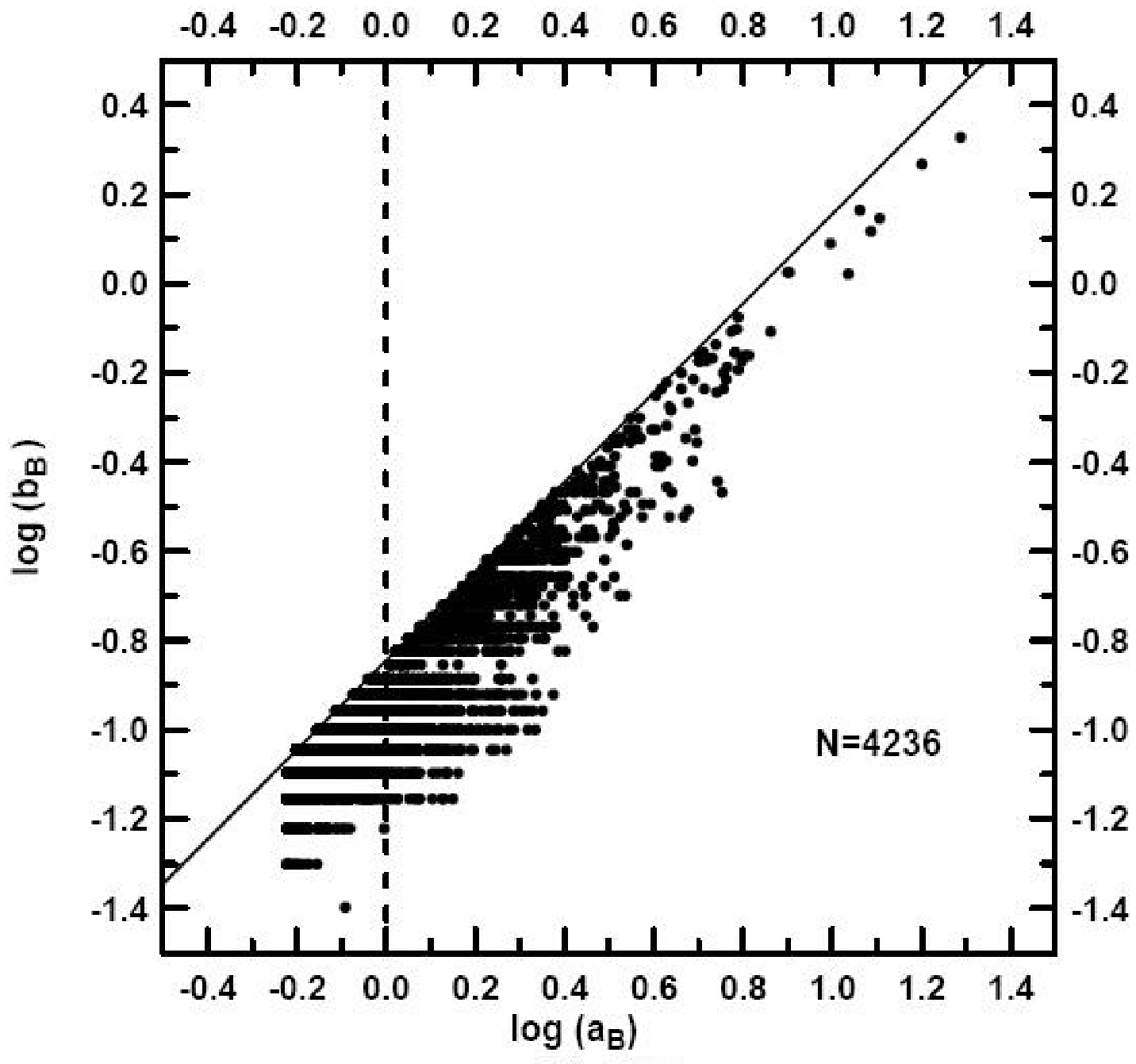}
\includegraphics[width=0.475\textwidth]{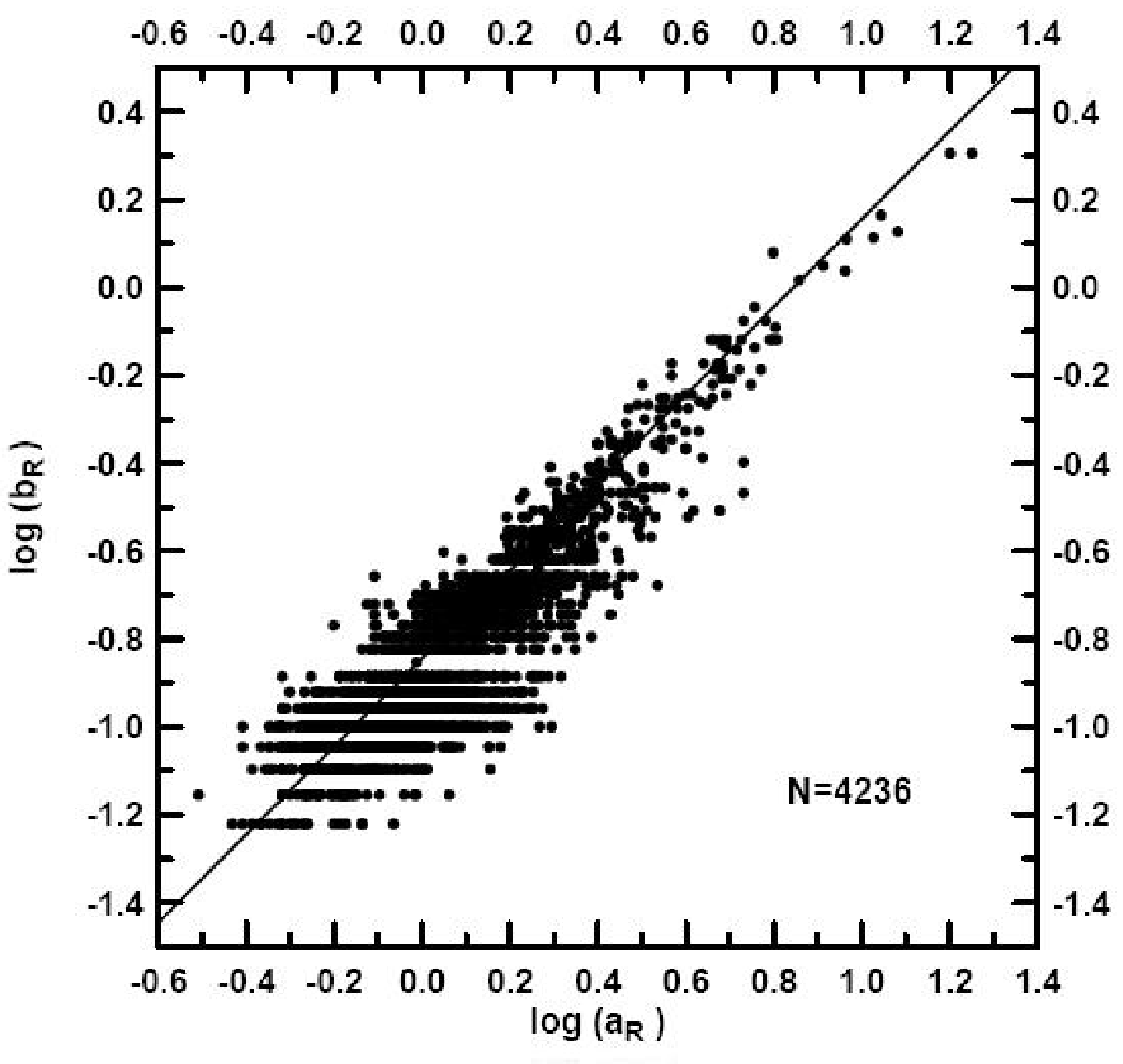}
\caption{The left plot shows the distribution of 4236 flat galaxies
of the RFGC catalog by the blue major and minor  diameters
$\log\,(b_B)$ vs. $\log\,(a_B)$ in minutes of arc. The thin line
corresponds to the condition $(a/b)_B=7$, i.e.
$\log\,(b_B)=\log\,(a_B) - 0.845$. The dashed line is drawn for the
value of  $a_B = 1\farcm0$. The right plot shows the distribution of
4236 RFGC flat galaxies
 by the red major and minor diameters   $\log\,(b_R)$ from
$\log\,(a_R)$. The thin line corresponds to the condition
$(a/b)_R=7$.}
\end{figure*}

\renewcommand{\baselinestretch}{0.9}
\begin{table*}
\setcaptionmargin{0mm} \onelinecaptionstrue \captionstyle{normal}
\caption{The distribution of RFGC  galaxies by the blue major
diameters and radial velocities $V_{\rm LG}$}
\medskip
\begin{tabular}{c|c|c|c|c|c|c}
 \hline
$V_{\rm LG}$, km\,s$^{-1}$ & $a_B\geq2\farcm0$ & $1\farcm99$--$1\farcm50$ & $1\farcm49$--$1\farcm00$ & $0\farcm99$--$0\farcm60$ & $a_B\geq0\farcm6$ & $a_B\geq1\farcm0$\\
\hline
(0--1000]               &4/35 (11)   & 2/11 (18)    & 0/5 (0)      & 0/5 (0)    & 6/56 (11)     & 6/51 (12)\\
(1000--2000]            &16/69 (23)  & 9/27 (33)    & 5/23 (22)    & 3/22 (14)  & 33/141 (23)   & 30/119 (25)\\
(2000--3000]            &20/79 (25)  & 7/36 (19)    & 7/57 (30)    & 3/28 (11)  & 47/200 (24)   & 44/172 (26)\\
(3000--4000]            &12/38 (32)  & 16/47 (34)   & 19/61 (31)   & 3/36 (8)   & 50/182 (27)   & 47/146 (32)\\
(4000--5000]            &14/39 (36)  & 14/65 (22)   & 39/125 (31)  & 8/53 (15)  & 75/282 (27)   & 67/229 (29)\\
(5000--6000]            &12/33 (36)  & 15/53 (28)   & 40/129 (31)  & 8/65 (12)  & 75/280 (27)   & 77/215 (36)\\
(6000--7000]            &9/24 (38)   & 14/36 (39)   & 41/117 (35)  & 6/108 (6)  & 70/285 (25)   & 64/177 (36)\\
(7000--8000]            &5/14 (36)   & 8/34 (24)    & 31/89 (38)   & 6/99 (6)   & 50/236 (21)   & 44/137 (32)\\
(8000--9000]            &3/6 (50)    & 12/26 (46)   & 30/90 (33)   & 4/92 (4)   & 49/214 (23)   & 45/122 (37)\\
(9000--10\,000]         &2/2 (100)   & 3/18 (17)    & 22/71 (31)   & 8/111 (7)  & 35/202 (17)   & 27/91 (30) \\
(0--10\,000]            &97/339 (29) & 100/353 (28) & 244/767 (32) & 49/620 (8) & 490/2078 (24) & 441/1459 (30)\\
$>$ 10\,000             &2           & 27           & 255          & 657        & 941           & 284\\
RFGC, with $V_{\rm LG}$ &341         & 380          & 1022         & 1277       & 3020          & 1743\\
All RFGC                &343         & 384          & 1174         & 2335       & 4236          & 1901\\
\hline
\end{tabular}
\end{table*}

\renewcommand{\baselinestretch}{1.0}

 The distribution of the RFGC galaxies by the blue ($B$) and red
($R$)
 axial ratios is demonstrated in Fig.~2. As shown
in~\cite{kar1997:Karachentseva_n}, the linear regression between them has the shape of $(a/b)_R = 0.853 (a/b)_B$  with a rather large dispersion. To sever the
galaxies with prominent bulges, we chose the following
conditions as a criterion for a super-thin
(ultra-flat) galaxy (Ultra Flat Galaxies = UFG):
 $(a/b)_B\geq10.0$ and $(a/b)_R\geq 8.53$ (the upper right corner of Fig.~2.)

Obviously,  the statistics of the observed properties of
ultra-thin galaxies selected this way is affected by various selection effects.
Let us list   the main ones:
\begin{list}{}{
\setlength\leftmargin{2mm} \setlength\topsep{2mm}
\setlength\parsep{0mm} \setlength\itemsep{2mm} }
 \item (1) In the remote/small/faint galaxies the axial ratio is determined,
basically, by the size of the minor axis $b$, which is affected by
the resolution of the photographic emulsion and the seeing. Thus, for
a galaxy with  $a =36\arcsec$ and $a/b=10$ the value amounts to
\linebreak $b=3\farcs6$,
 which is comparable to the typical resolution on the photographs of the Palomar sky survey of \linebreak about $3\arcsec$.
\item (2) The amount of data on the radial velocities of distant,  $V_{\rm LG}>10\,000$ km s$^{-1}$ galaxies rapidly decreases  with
increasing distance, hence, it becomes difficult to estimate the number of
their natural satellites.
 \item (3) The shapes of the galaxies in the zone of the Milky Way, especially the edge-on late spirals
 are affected by the absorption in our own Galaxy, as well as the foreground stars.
\end{list}


The angular size distribution of the RFGC galaxies and the influence
of the above-mentioned effects can well be seen on the panels of
Fig.~3, where the
 ``$\log\,(b_B)$--$\log\,(a_B)$'' and ``\mbox
{$\log\,(b_R)$--$\log\,(a_R)$}'' diagrams are presented to the left and right, respectively.
A thin line on both panels is determined by the condition $a/b=7$, i.e.
 $\log\,(b) = \log\,(a) - 0.845$.  The following features are distinctly
  conspicuous: \linebreak a) in  red  color the RFGC galaxies on the average look thicker than in blue color;
  b) the diameter dispersion  increases with  decreasing size of  galaxies;
c)  at the value of $\log\,(b) \sim -1.3$ $(b \sim 3\arcsec$)
  the shortage of the number of galaxies becomes noticeable caused by    the limit,
resulting from the resolution of the photographic emulsion. For
small-sized galaxies the measurement resolution of minor diameter is
evident.

A simultaneous fulfillment of the conditions\linebreak \mbox
{$(a/b)_B\geq10.0$} and $(a/b)_R\geq8.53$ isolates 19\% of ultra-flat
galaxies with the diameter of  \mbox {$a_B\geq0\farcm6$} from the
entire RFGC catalog. We shall designate this sample of 817 galaxies
the ``base UFG.'' Taking into account the above-mentioned selection
effect we introduce additional restrictions:  \mbox {$V_{\rm LG} \leq
10\,000$~km\,~s$^{-1}$}, and galactic latitude \mbox {$|b| >
10\degr$}. Four hundred and ninety  objects satisfy the   required
criteria of ultra-thin galaxies   (the \mbox{``$N = 490$''} sample).


Table~1 gives a comparison of the two-dimensional distributions of
the RFGC galaxies by the radial velocities in the Local Group system
and the major blue angular diameters. We have adopted the radial
velocity estimates from the NED \footnote{{\tt
www.ned.ipac.caltech.edu}} and HyperLeda\footnote{{\tt
http://leda.univ-lyon1.fr/}} databases. The denominator contains the
number of all the RFGC galaxies
 $(a_B\geq0\farcm6$,
$(a/b)_B\geq7$) with $V_{\rm LG} \leq 10\,000$ km s$^{-1}$ (this is
the ``$N = 2078$'' sample),  the numerator contains the number of all
the ultra-thin galaxies, while the percentage of ultra-thin galaxies
in the appropriate bin is in the brackets.


The data of Table~1 demonstrate a significant decrease in the
fraction of ultra-thin galaxies with the angular diameters from
$0\farcm99$ to $0\farcm60$ for all the    radial velocity ranges. No
significant difference is observed in the fraction of ultra-thin $N =
490$ sample galaxies  in all the    radial velocity ranges, except
the first (with a small number of objects) and the last one. Recall
that the statistics for the distant galaxies is affected by the
selection effects noted above. Therefore, to achieve an acceptable
completeness we select a yet more refined sample,  $N = 441$ (the
last column of Table~1) in which the  $a_B\geq1\farcm0$ condition is
fulfilled instead of  $a_B \geq 0\farcm6$.
 It amounts to about 10\% of the total number of flat RFGC galaxies.
Earlier, in~\cite{kud1997:Karachentseva_n}  we found that the FGC(E) catalog itself is about 90\% complete at  $a_B\geq1\farcm0$.
 The data of Table~1 shows that  the sample of ultra-thin  galaxies is as well almost complete for the galaxies with the velocities of less than
 10\,000~km\,s$^{-1}$ exactly given $a_B\geq1\farcm0$.


The Schmidt  test $\langle
V/V_{max}\rangle$~\cite{sch1968:Karachentseva_n} for the \mbox {$N =
441$} sample is given in Fig.~4. For the maximum and initial angular
diameter   the values of $6\arcmin$ and $0\farcm6$ have been taken,
respectively. It is evident that   the completeness at the level of
80--90\% is achieved at \mbox {$a_{\rm min} \sim 1\farcm0$}.


\begin{figure}
\setcaptionmargin{5mm} \onelinecaptionstrue \captionstyle{normal}
\includegraphics[width=0.95\columnwidth,]{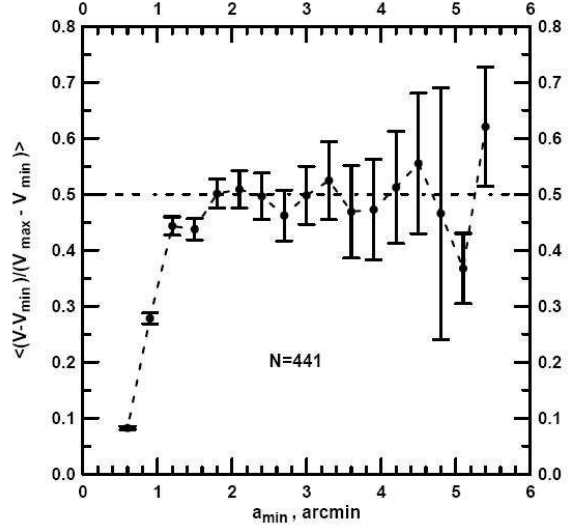}
\caption{The Schmidt  test for the $N=441$ sample}
\end{figure}

The Annex gives a list of RFGC numbers for 817 galaxies of the base
\mbox{UFG-sample}. The RFGC-numbers of galaxies of the refined
\mbox{UFG-sample} \mbox{$N = 441$} are marked with two asterisks, 49
galaxies outside the UFG-sample, but included in the $N = 490$
sample, are marked with one asterisk.


Figure~5 gives the maps of sky distributions in the equatorial
coordinates for the RFGC galaxies (points) and the UFG sample (filled
circles). The gray watercolour wash designates the region of strong
absorption near the Galactic equator  $\mid b\mid \leq 10\degr$. The
following sections by the radial velocities are presented:  $V_{\rm
LG} < 3000$~km\,s$^{-1}$, \mbox {$3000< V_{\rm LG} <
10\,000$~km\,s$^{-1}$}, $V_{\rm LG}> 10\,000$~km\,s$^{-1}$, the
radial velocities are not measured. The array of the images gives an
idea of the relative positions  of the UFG sample objects and all the
flat   galaxies  from the RFGC catalog at different depths.


On the top panel of Fig.~5 we can see that the nearby ultra-thin  galaxies
barely outline the Local Supercluster. An excess of UFGs in the region of the
    Local Supercluster center  (\mbox {${\rm RA} = 12\fh5$}, \mbox
{${\rm Dec} = +12\degr$})  as compared to the homogeneous distribution
 is only  \mbox {$\Delta N \sim5$} of galaxies. The section of \mbox
{3000--10\,000}\,km\,s$^{-1}$  (the second panel from the top) is
filled the most complete mainly due to the observations with the
300-m Arecibo  Observatory radio
telescope~\cite{gio1997:Karachentseva_n}, the 6-m BTA
telescope~\cite{mak2001:Karachentseva_n} and the 100-m Radio
Telescope
Effelsberg~\cite{huc2005:Karachentseva_n,mit2005:Karachentseva_n}.
The galaxies in this radial velocity interval also show a barely
discernible concentration in the regions of the well-known    Coma
and \mbox{Pisces--Perseus} clusters. Note that Fig.~5 and the data in
Table~1 are highly complementary. This way,  the two last  lines of
Table~1 for the galaxies with the angular diameters in the range of
$0\farcm60$--$0\farcm99$ show a noticeable lack of radial velocity
measurements. Wherein an excessive localization of galaxies without
radial velocities is observed in the southern hemisphere (the bottom
panel).

\begin{figure}
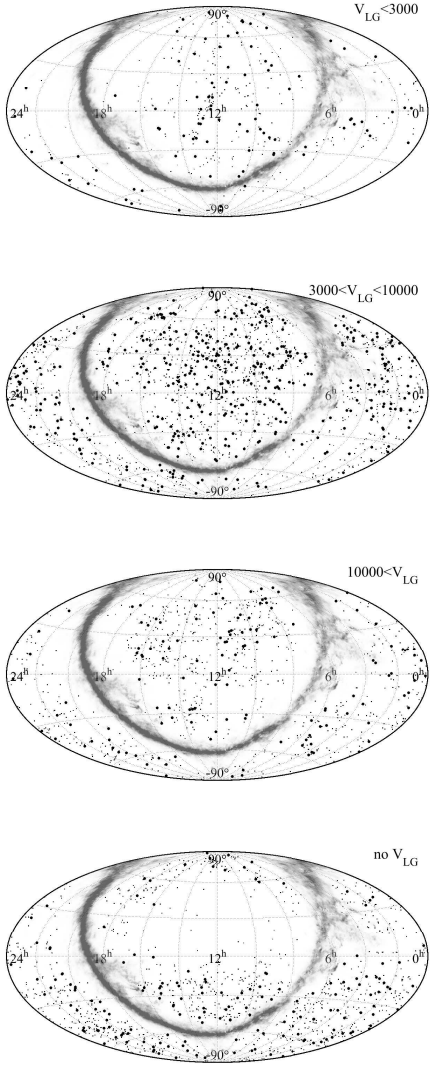

\setcaptionmargin{5mm} \onelinecaptionstrue \captionstyle{normal}
\includegraphics[width=0.45\textwidth, clip]{Karachentseva_fig5a.eps}
\includegraphics[width=0.45\textwidth, clip]{Karachentseva_fig5b.eps}
\includegraphics[width=0.45\textwidth, clip]{Karachentseva_fig5c.eps}
\includegraphics[width=0.45\textwidth, clip]{Karachentseva_fig5d.eps}
\caption{The  maps of the sky distribution in the equatorial
coordinates of the RFGC galaxies (points) and the UFG sample (filled
circles). Gray watercolor wash designates the region of strong
absorption near the Galactic equator  $\mid b\mid \leq 10\degr$. From
top to bottom, the following sections by radial velocities are
presented: \mbox {$V_{\rm LG}< 3000$}~km\,s$^{-1}$, $3000< V_{\rm LG}
< 10\,000$~km\,s$^{-1}$, $V_{\rm LG}> 10\,000$~km\,s$^{-1}$, and
radial velocities are not measured.}
\end{figure}

\section{THE PROPERTIES OF ULTRA-THIN GALAXIES COMPARED
WITH THE RFGC GALAXIES}

Among the 2078 galaxies of the RFGC catalog   with radial velocities
$V_{\rm LG} <10\,000$~km\,s$^{-1}$, the average radial velocity in
the  Local Group rest frame amounts to  $5553\pm54$~km\,s$^{-1}$. For
the UFG sample of galaxies with angular diameters exceeding
$0\farcm6$ ($N = 490$) and  $1\farcm0$ ($N = 441$), the average
velocity value is $5438\pm104$~km\,s$^{-1}$ and
\mbox{$5366\pm110$}~km\,s$^{-1}$, respectively. This means that all
the three samples   insignificantly differ in depth.


Table~2 shows the distribution of galaxies by the
morphological types of spirals in the initial RFGC catalog, in the sample of
ultra-thin galaxies, and in the   RFGC  and UFG samples with measured
velocities. The last two rows of Table~2 give the percentage of   the  ultra-thin galaxy sample objects in the corresponding bins.
The first line contains the designations of the  spiral types in the
Hubble system, the second line---in the de~Vaucouleurs digital system.
As it was shown in~\cite{kar1997:Karachentseva_n}, our estimates of different types
vary from the digital estimates of de Vaucouleurs on the average by not more than
 $\pm 1$.
In the same paper~\cite{kar1997:Karachentseva_n} it was noted that
the RFGC galaxies show no dependence of the apparent axis ratio (or
type) on the radial velocity up to values of about
10\,000--12\,000~km\,s$^{-1}$. As follows from the  Table~2 data, the
peak  number of galaxies for the entire catalog, as well as the $N =
2078$ sample  falls on   the Sc type. For the  UFG $N = 817$ and $N =
441$  samples the peak shifts to the  Sd type, both in the general
and in the fraction distributions. The fraction of ultra-thin
galaxies  rapidly decreases from the  Sd type to later Sdm  and Sm
types, since   turbulent motions play a significant role in the
dynamics of the latter.
 As we can see,  the morphological types  of 80\% of \mbox{UFG-galaxies}
lie in the narrow range of   $T = 7\pm1$.
This is consistent with the result of~\cite{hei1972:Karachentseva_n}: the  thinnest
 stellar disks are found in the galaxies classified as Scd,
Sd, Sdm.

\begin{table*}
\setcaptionmargin{0mm} \onelinecaptionstrue \captionstyle{normal}
\caption{The distribution of RFGC galaxies and ultra-flat galaxies by the type of spirals}
\medskip
\begin{tabular}{l|c|c|c|c|c|c|c|c|c}
 \hline
\multicolumn{1}{c|}{Type} & Sab & Sb & Sbc & Sc & Scd & Sd & Sdm & Sm & All\\
 \hline
 \multicolumn{1}{c|}{T}   & 2   & 3  & 4   & 5  & 6   & 7  & 8   & 9  &--\\
 \hline
RFGC           & 1  & 151 & 573 & 1535 & 960 & 718 & 252 & 37 & 4236\\
$N=2078$       & 8  & 82  & 266 & 544  & 490 & 465 & 195 & 28 & 2078\\
UFG, $N = 817$ & 0  & 11  & 31  & 252  & 211 & 270 & 40  & 2  & 817\\
UFG, $N = 441$ & 0  & 9   & 17  & 81   & 109 & 188 & 35  & 2  & 441\\
N817/NRFGC,\% & 7  & 5   & 16  & 22   & 38  & 16  & 5   & 19 & \\
N441/N2078,\% & 11 & 6   & 15  & 22   & 40  & 18  & 7   & 21 & \\
 \hline
\end{tabular}
\end{table*}

\begin{table}
\setcaptionmargin{0mm} \onelinecaptionstrue \captionstyle{normal}
\caption{The distribution of RFGC galaxies and super-thin galaxies by the
mean surface brightness index, SB}
\medskip
\begin{tabular}{l|c|c|c|c|c}
 \hline
\multicolumn{1}{c|}{SB} & I & II & III & IV & All\\
 \hline
RFGC           & 242 & 2480 & 1369 & 145 & 4236\\
$N = 2078$     & 190 & 1306 & 534  & 48  & 2078\\
UFG, $N = 817$ & 23  & 451  & 310  & 33  & 817\\
UFG, $N = 441$ & 22  & 268  & 139  & 12  & 441\\
N817/NRFGC,\% & 10  & 18   & 23   & 23  & 19\\
N441/N2078,\% & 12  & 20   & 26   & 25  & 21\\
 \hline
\end{tabular}
\end{table}

\begin{table}
\setcaptionmargin{0mm} \onelinecaptionstrue \captionstyle{normal}
\caption{The distribution of RFGC galaxies and super-thin  galaxies by the
asymmetry index, As}
\medskip
\begin{tabular}{l|c|c|c|c}
 \hline
\multicolumn{1}{c|}{As} & 0 & 1 & 2 & All\\
 \hline
RFGC           & 2830 & 1159 & 247 & 4236\\
$N = 2078$     & 1260 & 640  & 178 & 2078\\
UFG, $N = 817$ & 568  & 209  & 40  & 817\\
UFG, $N = 441$ & 272  & 135  & 34  & 441\\
N817/NRFGC,\% & 20   & 18   & 16  & 19\\
N441/N2078,\% & 22   & 21   & 19  & 21\\
\hline
\end{tabular}
\end{table}

The surface brightnesses for the RFGC galaxies were estimated
visually and were divided by  the average surface brightness index
SB: I, II, III and IV (where the class I galaxies have the highest
surface brightness). Table~3 shows the distribution of the number of
flat RFGC galaxies,  the $N = 2078$ and $N = 441$  samples by the
surface brightness class SB.  The maximum in the distributions of
55--63\% for all the samples falls on \mbox{${\rm SB} = {\rm II}$}
(what corresponds to about 25.4~mag/arcsec$^2$ in the
$B$-band~\cite{kar1993:Karachentseva_n}). The last two rows of
Table~3 specify the percentage of UFG galaxies  among the
RFGC-galaxies of each SB class  with and without  radial velocity
measurements.  These data show that there has been a decrease in the
fraction of high surface brightness  objects in the transition from
flat RFGC-galaxies to ultra-flat UFGs. This trend corresponds to the
expected dropout of galaxies with small bulges (Sbc type) compiling
the UFG sample. In other words, in the transition from flat to
ultra-flat galaxies there is a shift of the SB index towards faint
surface brightness, ${\rm SB} = {\rm III, IV}$ (or about
 25.6--25.9~mag/arcsec$^2$~\cite{kar1993:Karachentseva_n}).


Thus, a typical super-thin galaxy is a Sd-type spiral galaxy with low
surface brightness. The diminished surface brightness can be caused
by a stronger absorption of light in the ultra-thin galaxies seen
strictly edge-on, or rather by the lower density of their stellar
disks.


The asymmetry of shape of a flat galaxy is quite difficult to estimate without
a deep enough surface photometry. Hence the results
given in Table~4  should be evaluated as preliminary.
The shape asymmetry was characterized in the RFGC catalog as the As index,
which took the values  0, 1 and 2 for the regular, intermediate and
clearly disturbed forms, respectively. As we can see from the table data, flat galaxies of regular shapes  make up from 62 to 70\% and the most asymmetric---from 5 to 8\% both throughout the catalog, and in the individual samples.
It also  follows from the data in Table~4  that the proportion of ultra-thin galaxies of different types of asymmetry
among the $N = 2078$ sample of galaxies  is about 20\%,
showing a slight tendency to decrease from  regular   to
intermediate and disturbed  shapes.


Therefore, our data shows no distinct relationship between the
relative thickness of the stellar disk of a spiral galaxy and the
degree of perturbations of its periphery.

\section{THE PROPERTIES OF ULTRA-THIN GALAXIES
DEPENDING ON THEIR ENVIRONMENT}

We determined the density of the environment in several ways.

(1)  In the RFGC catalog for each galaxy with the $a_B$ diameter
 significant neighbors with angular diameters in the range of
[$a_B/2 - 2\,a_B$], located in a circle of $R=10\,a_B$ radius were counted.
 At that, the neighbors were identified  the same way for all the galaxies.
By the time of the catalog publication, the data on the radial
velocities of RFGC galaxies and especially their fainter neighbors
were very scarce. Therefore, the given  numbers of neighbors in the
projection  give an idea only on the surface density of the
background  around the RFGC galaxies not accounting for  its depth.
Figure~6 demonstrates the histograms of  distribution of the galaxies
from four samples:  RFGC, $N = 2078$, $N = 490$ and $N = 441$ by  the
number of significant neighbors.  The right-hand side scale in each
panel indicates the percentage of galaxies in the corresponding
sample with the designated number of neighbors in the projection. The
behavior of the distribution is about the same   for the considered
samples, and  only 2--5\% of the galaxies in each sample and in the
entire RFGC catalog have more than three neighbors. As follows from
the data in Fig.~6, there is a slight upward trend in the proportion
of isolated galaxies in the transition from  flat to  ultra-flat
galaxies.

 (2) For each  super-thin galaxy of the UFG samples we determined
the  number of galaxies with relative radial velocities in the range of
\mbox{$+500$}, $-500$~km\,s$^{-1}$ to the limiting projection
distance of $R = 750$~kpc. Unlike in the previous case,
 radial velocities of galaxies were taken into account here. However,
 the neighbors, as in selection (1) can not reliably form physical
systems with the UFG sample galaxies, although they share with them a
fairly well-indicated  general field by the velocities and
distances.


The distribution of ultra-thin galaxies by the number of such
neighbors for the   $N = 490$  and $N = 441$ samples is shown in
panels of Fig.~7 to the left and right, respectively. The left-hand
side scale in the panels shows the number of galaxies in the bin,
while  the right-hand side scale indicates their percentage. The last
value on both panels corresponds to the cases of seventeen or more
neighbors.

\begin{figure*}
\setcaptionmargin{5mm} \onelinecaptionstrue \captionstyle{normal}
\includegraphics[width=\columnwidth]{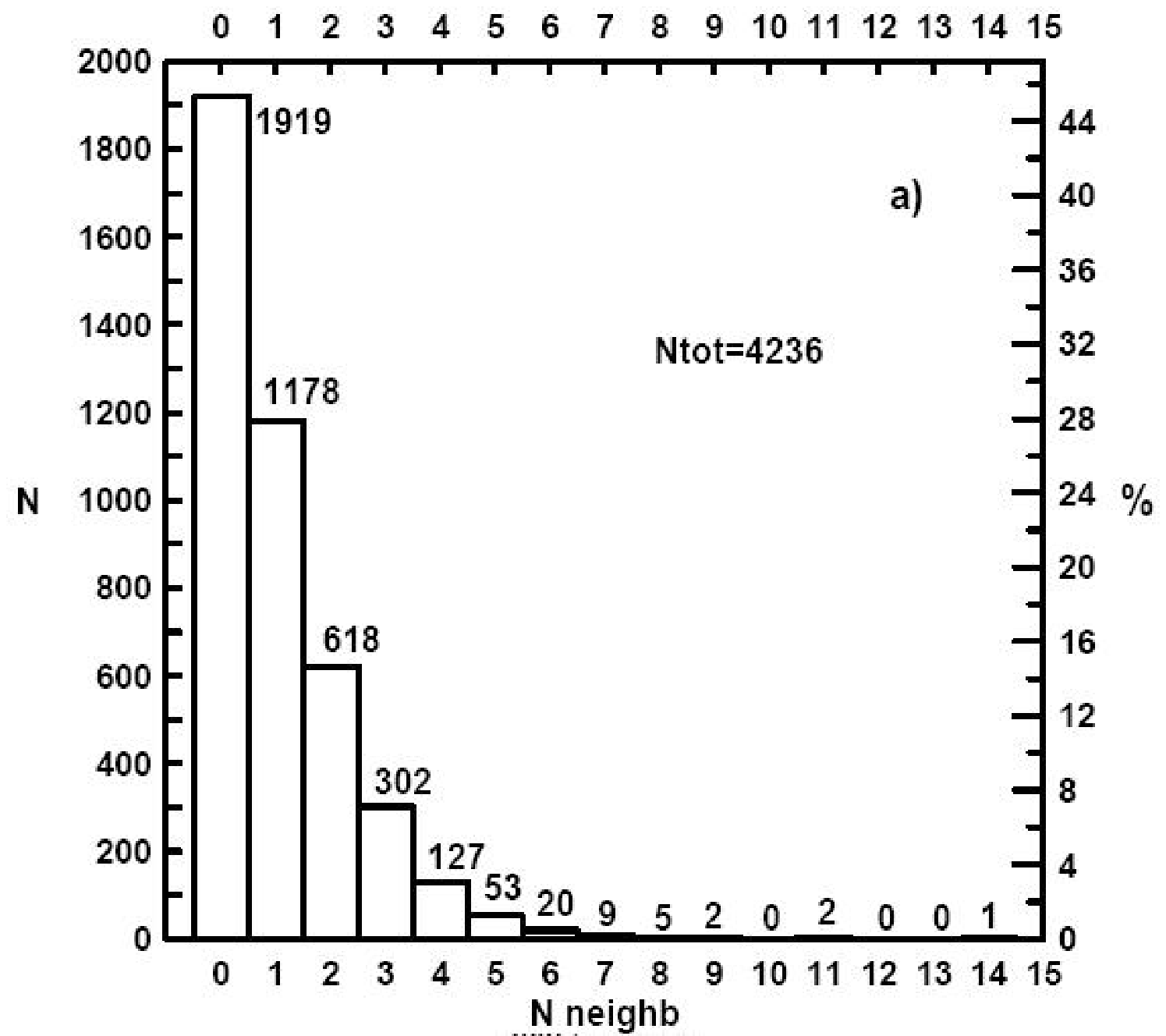}
\includegraphics[width=\columnwidth]{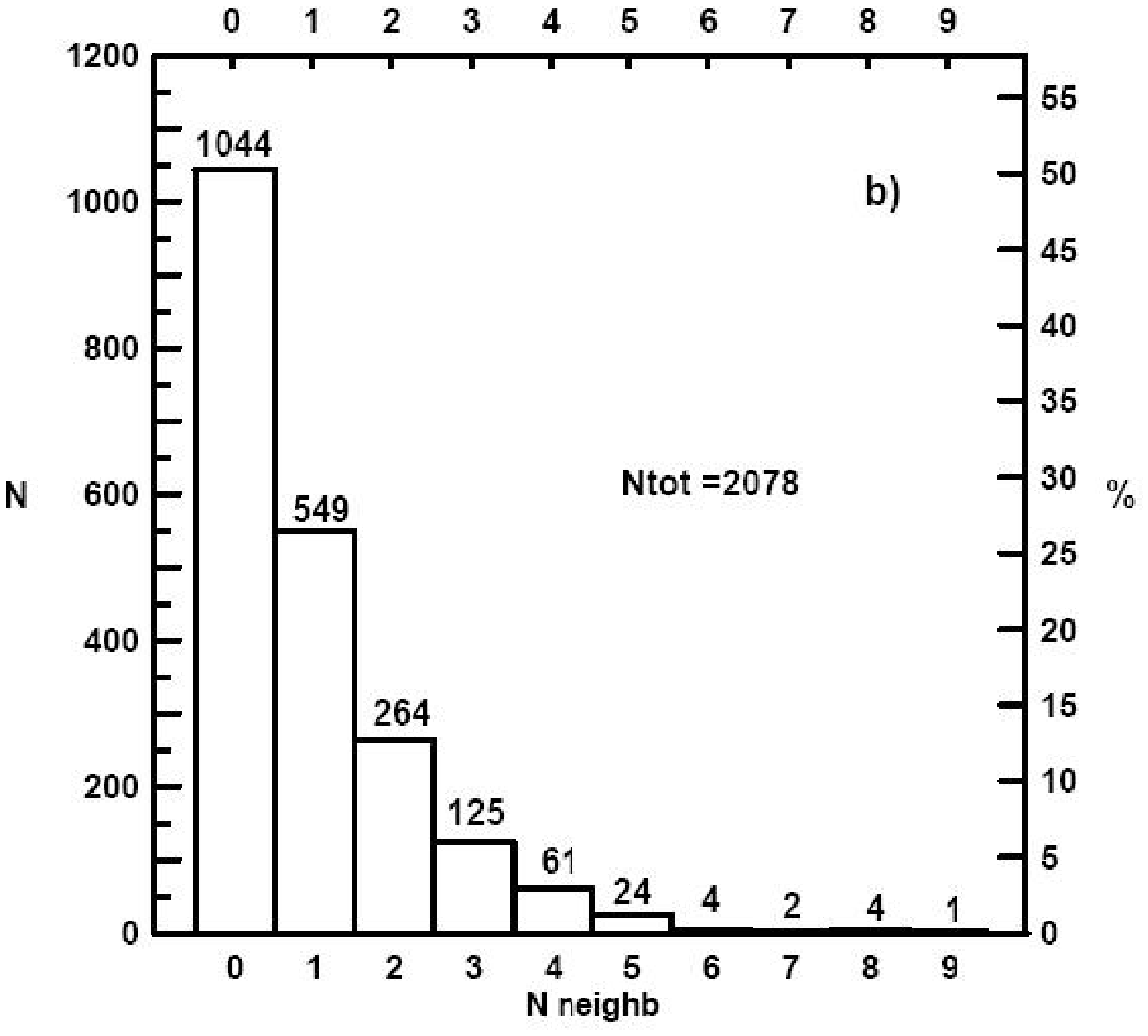}
\includegraphics[width=\columnwidth]{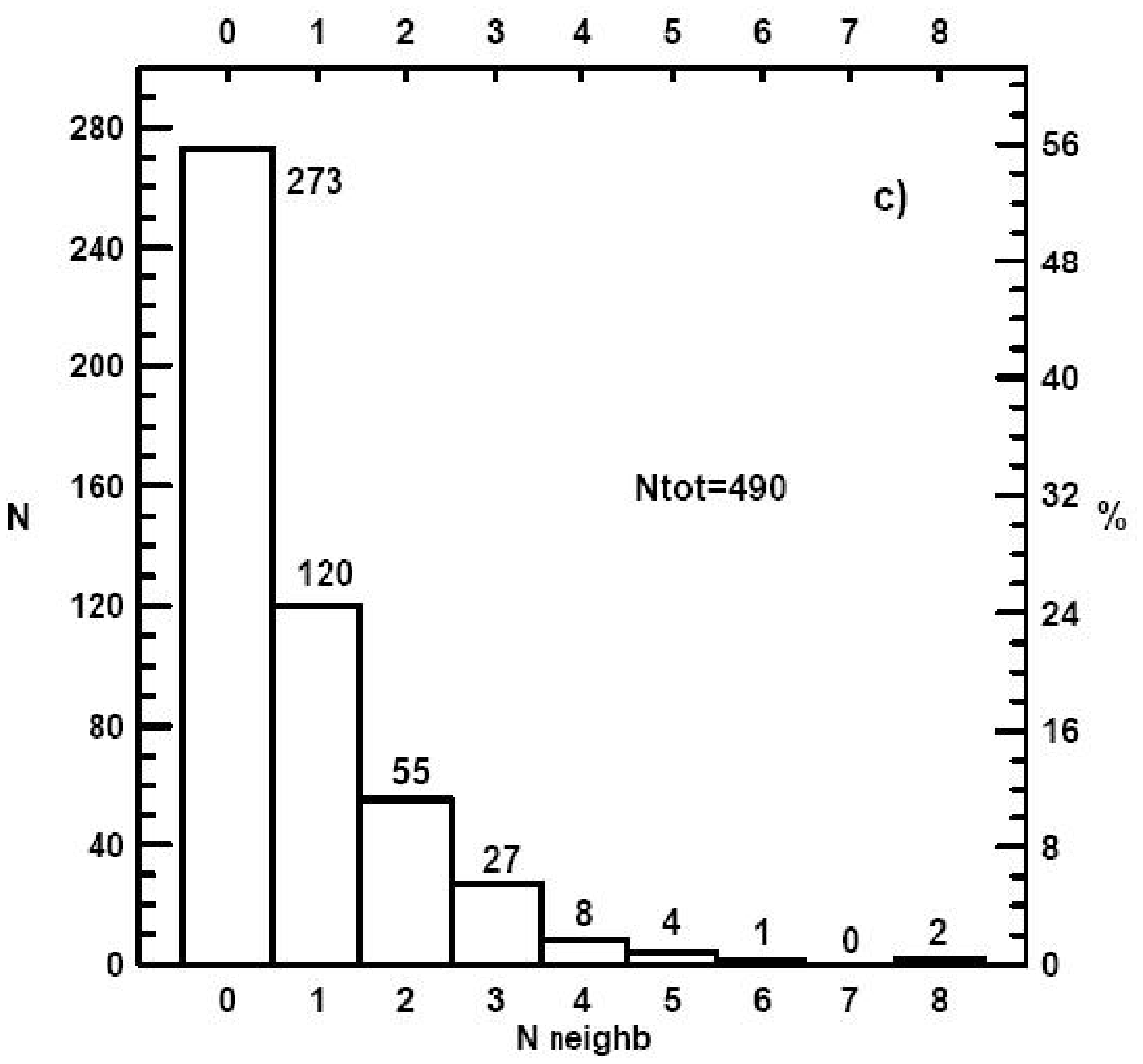}
\includegraphics[width=\columnwidth]{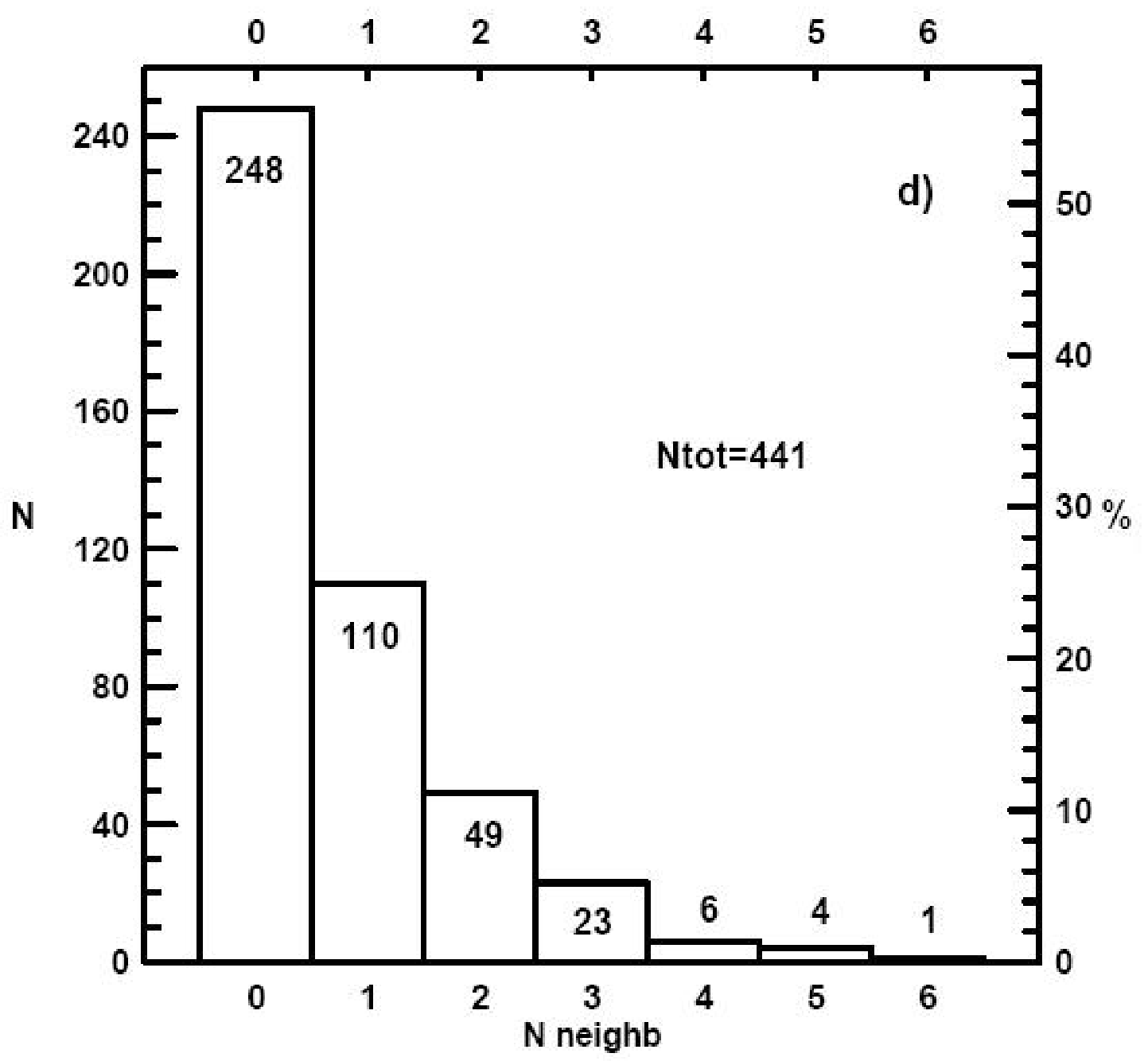}
\caption{The distribution of flat galaxies by the number of
significant neighbors with angular diameters in the range of [$a_B/2
- 2\,a_B$]  in a circle of $R=10\,a_B$ radius  for: (a) the entire
RFGC catalog, (b) the RFGC sample with radial velocities below
10\,000~km\,s$^{-1}$, $N = 2078$, (c) sample of ultra-thin galaxies
with $a_B\geq 0\farcm6$, $N = 490$, (d) sample of ultra-thin galaxies
with $a_B\geq 1\farcm0$.}
\end{figure*}

We can see from the data of Fig.~7 that only a third (31\%) of ultra-thin
galaxies   have no neighbors within this range of radial velocities
and projection distances. This value is smaller than the proportion of
isolated galaxies, 56\%, taking the neighbors into account by method~(1).
It should be noted, however, that not all the neighbors with the velocities
in the range of $+500$, $-500$~km\,s$^{-1}$ and projection distances
of \mbox{$R<750$}~kpc are the natural companions of ultra-flat
galaxies. Some of them may belong together with the UFGs to the
diffuse elements of the large-scale structure (filaments  and walls).

\begin{figure*}
\setcaptionmargin{5mm} \onelinecaptionstrue \captionstyle{normal}
\includegraphics[width=\columnwidth]{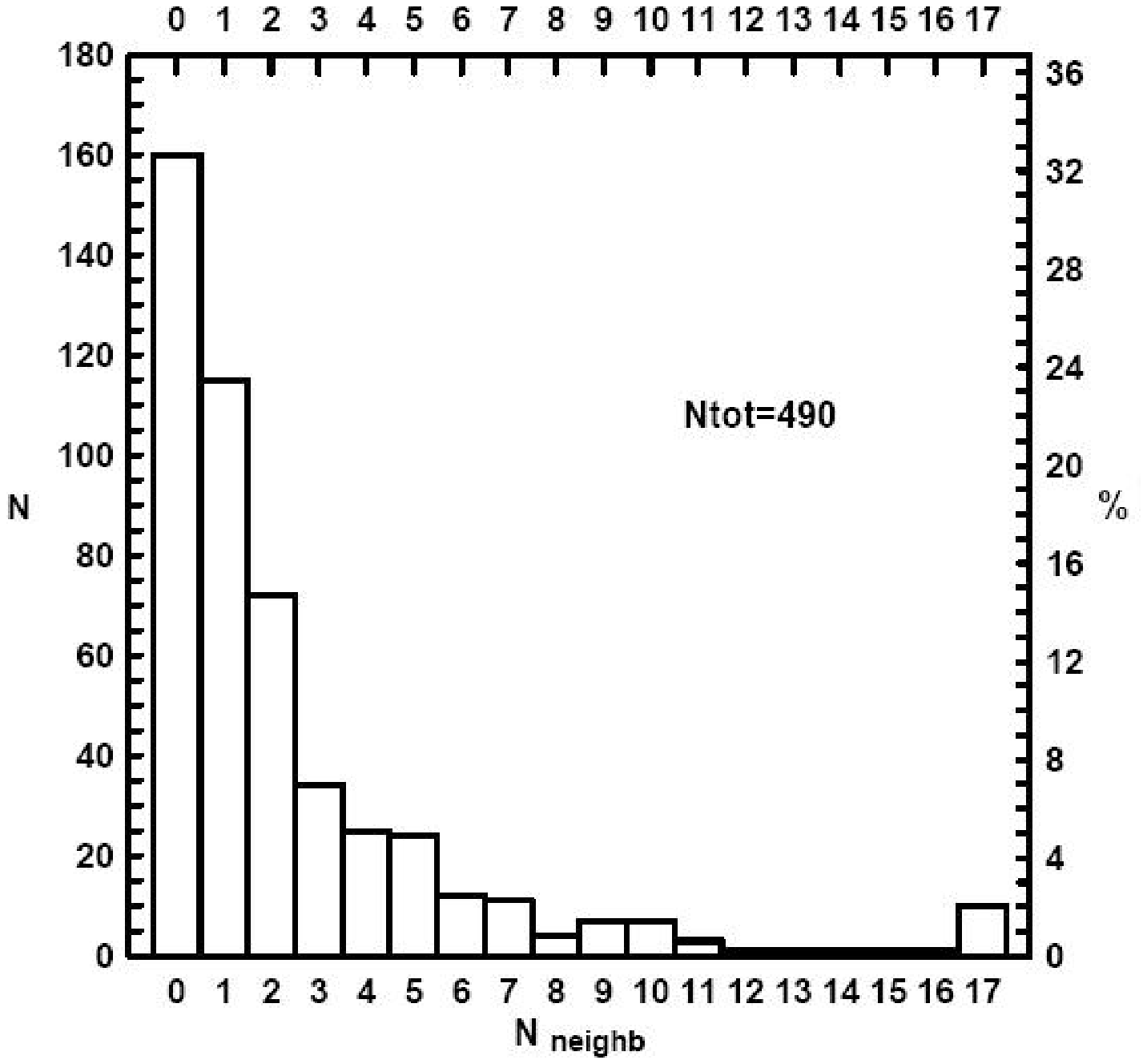}
\includegraphics[width=\columnwidth]{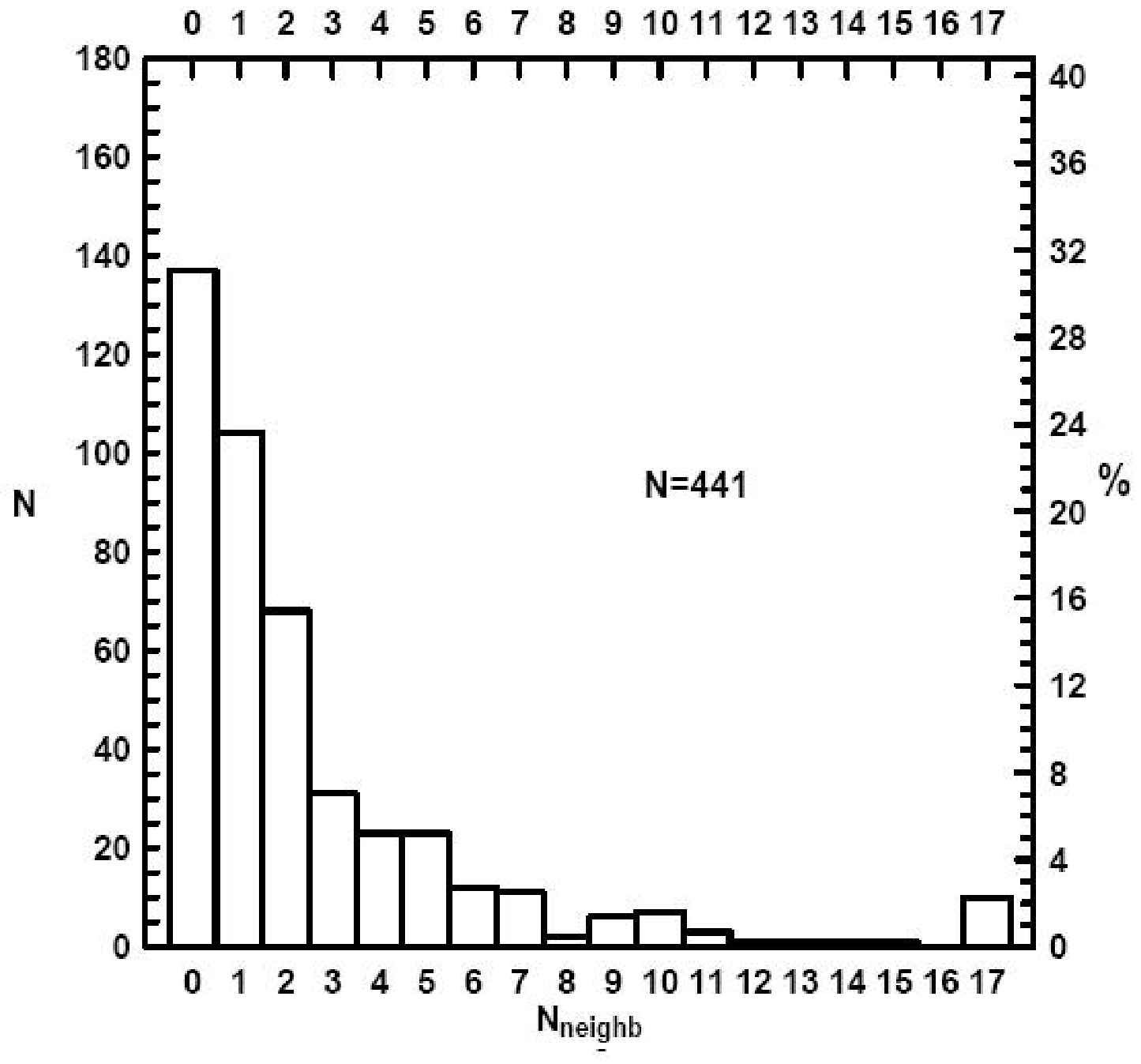}
\caption{The distribution of ultra-thin galaxies by the number of neighbors
up to the limiting projection distance of \mbox {$R = 750$}~kpc and with the
radial velocities of $V_{\rm LG}$  over the range of $+500$,
$-500$~km\,s$^{-1}$; the $N = 490$ sample (left), the  $N = 441$ sample
(right).}
\end{figure*}

 (3) To all the galaxies with radial velocities below 20\,000~km\,s$^{-1}$ in the
HyperLeda database at  the galactic  latitude of  $\mid
b\mid>10\degr$  the clustering algorithm, described in detail
in~\mbox{\cite{kar2008:Karachentseva_n,mak2011:Karachentseva_n}} was
applied. Uniting the galaxies into  systems of different populations,
individual characteristics of all the galaxies were used, namely,
their radial velocities, projected separations and masses determined
by the integral infrared luminosity in the infrared
\mbox{$K_s$-band}. Initially, we isolated physical pairs satisfying
the conditions of the full negative energy, and pair location within
the ``zero velocity'' sphere~\cite{san1986:Karachentseva_n}. Then all
the pairs with any common component were combined into a group.
Eventually we had the data for the entire sky on the groups of
galaxies and their environment up to 10\,000~km\,s$^{-1}$. The
resulting catalog was used for the analysis of the environment of
galaxies of the RFGC catalog and the $N = 490$ and $N = 441$ samples.
In this paper we used the following statuses to describe the
environment: ``isol'' for isolated galaxies, ``root'' for the main
member of the group, ``mem'' for a member of the group. Hence,
method~(3) gives a more accurate representation of the physical
environment of ultra-thin galaxies than method~(2), and the more so
method~(1).


For the $N  = 490$ and $N = 441$  samples, we determined the
frequency of galaxy occurrence in different environments. We
considered the cases of selection by method~(3) with the ``isol,''
``root'' and ``mem'' status of a galaxy, as well as by method~(2),
when a galaxy has 0, 1, and 2 or more neighbors. The results are
presented in Table~5. As follows from these data, more than 60\% of
ultra-thin galaxies belong to the category of dynamically isolated
objects, about 30\% are nonprincipal members of scattered groups
(associations, filaments or walls) and only the tenth of them are
classified as dynamically dominant objects with respect to their
nearest neighbors. The numbers in three right-hand side columns of
Table~5 are consistent with the data of Fig.~6 showing that more than
a half of UFG galaxies, which have no neighbors by the selection
method~(2) are in fact dynamically isolated objects.

\begin{table}
\setcaptionmargin{0mm} \onelinecaptionstrue \captionstyle{normal}
\caption{The occurence of super-thin galaxies in different dynamic
environments}
\begin{tabular}{c|c|c|c|c|c|c}\hline
Number  & isol & root & mem &\multicolumn{3}{c}{neighbors} \\
 \hline
           &      &      &     & 0 & 1 & $\geq$2\\
 \hline
$N=490$    & 303      &  43     & 144      & 273      & 120      & 97\\
\%         & $62\pm4$ & $9\pm2$ & $29\pm3$ & $56\pm4$ & $24\pm3$ & $20\pm2$\\
$N=441$    & 267      & 41      & 133      & 248      & 110      & 83\\
\%         & $61\pm4$ & $9\pm2$ & $30\pm3$ & $56\pm4$ & $25\pm3$ & $19\pm2$\\
\hline
\end{tabular}
\end{table}

Analyzing the average characteristics of ultra-flat galaxies relating
to the categories ``isol,'' ``root'' and ``mem,'' we have noted the
following trends. The isolated ultra-thin galaxies on the average
have a later morphological type. The dimmest average surface
brightness is typical of isolated UFG galaxies. The degree of
asymmetry of an ultra-flat galaxy reveals a positive correlation with
the number of its neighbors.

\section{DISCUSSION AND SUMMARY}

Among the 4236 flat galaxies of the RFGC
catalog~\cite{kar1999:Karachentseva_n}, in which, by definition, the
blue axial ratios  satisfy the relation $(a/b)_B\geq7.0$, we have
selected a sample of ultra-flat spiral galaxies, Ultra Flat Galaxies,
UFG. This sample covers the entire northern and southern sky (except
for the zone of the Milky Way) and comprises 817 galaxies with blue
and red axial ratios \mbox{$(a/b)_B\geq10.0$} and \mbox
{$(a/b)_R\geq8.53$}, respectively. Within this initial (base) sample
of ultra-thin spirals seen edge-on, we fix a refined sample of 441
UFG galaxies, satisfying the following conditions: radial velocity of
the galaxy $V_{\rm LG}<10\,000$~km\,s$^{-1}$, Galactic latitude \mbox
{$\mid b\mid>10\degr$}, the major angular diameter
\mbox{$a_B\geq1\farcm0$}. According to the Schmidt test, a list of
441 UFG galaxies, designated as (**) in the Annex, is approximately
80--90\% complete and can serve as a model sample for the analysis of
various characteristics of ultra-flat galaxies.


We notice \mbox{UGC\,7321=RFGC\,2246} as a prototype of UFG-galaxies.
This is an isolated spiral with a blue and red axial ratio of 16 and
13 respectively. On the H$\alpha$-line image
from~\cite{kar2015:Karachentseva_n} a subsystem of young (about
10~Myr) stars,   immersed in the  H\,II regions, has an apparent
axial ratio of $(a/b)_{{\rm H}\alpha}  = 38$,  which is the highest
value among the flat galaxies. It is known that ultra-thin disks of
galaxies are mainly found  in the low number-density regions of the
surrounding galaxies. We have estimated the density of the UFG
environment in three different ways.
\begin{list}{}{
\setlength\leftmargin{2mm} \setlength\topsep{2mm}
\setlength\parsep{0mm} \setlength\itemsep{2mm} }

 \item (1) In the RFGC  catalog for each galaxy with the $a_B$ diameter,
 significant neighbors with the angular diameters in the range
of $a_B/2$--$2\,a_B$, located in a circle of   $R=10\,a_B$ radius were counted.
 \item (2) The number of galaxies with relative radial
velocities in the range of  $+500$, $-500$~km\,s$^{-1}$ to
the projected linear distance of $R = 750$~kpc was determined.
 \item (3) The environment density   was determined by the clustering method.
 \end{list}

A comparison of the UFG list  galaxies with the objects of the entire
RFGC catalog allows us to draw the following conclusions.
\begin{list}{}{
\setlength\leftmargin{2mm} \setlength\topsep{2mm}
\setlength\parsep{0mm} \setlength\itemsep{2mm} }

\item (a) More than 3/4 of UFG galaxies have their morphological
types in a narrow
 range of values   \mbox {$T= 7\pm1$}. In other words, the
thinnest stellar disks are found in the galaxies that are classified
as Scd, Sd, Sdm. This conclusion is in direct correspondence with the
previous results of Heidmann et al.~\cite{hei1972:Karachentseva_n}.

 \item (b) There appears a trend of diminishing average surface brightness
 in ultra-thin  galaxies in the transition   from the RFGC to the UFG sample.
It is mainly due to the elution of the Sbc-type galaxies with small
bulges from the UFG sample. The other probable reason could be a
stronger internal absorption of light in the ultra-thin galaxies seen
edge-on.

 \item (c) Regularly shaped disks of galaxies with no signs of asymmetry
(disturbances) make up about 2/3 both in the main RFGC catalog and in
the UFG sample. We observe only a slight tendency to an increase in
the relative number of undisturbed shapes in  super-thin galaxies. In
general, the thickness of the stellar disk of a spiral galaxy is not
related to the degree of disturbance (asymmetry) of its periphery.

 \item (d) Using different methods of estimating the surrounding density
of the UFG and RFGC galaxies, we have shown that the relative number
of isolated galaxies depends  on the apparent axis ratio of the
stellar disk only marginally.  According to our preliminary
estimates, about 60\% of ultra-flat galaxies can be classified as
dynamically isolated objects, about 30\% of them are part of the
scattered associations (filaments and walls), and only about 10\% of
them are dynamically dominant object relative to their nearest
neighbors.
\end{list}

In the following publications of this series we suggest to consider
the integral (optical and radio) properties of ultra-flat spiral
galaxies, and make the mass estimates of the dark haloes, enveloping
the disks of UFG galaxies.

\begin{acknowledgements}
The NED and HyperLeda databases were used in the study. IDK, DIM are
grateful to the Russian Science Foundation (grant 14--12--00965) for
the financial support of the study.
\end{acknowledgements}

 \onecolumngrid
\section*{ANNEX}

The list of RFGC numbers for 817 galaxies of the base sample. The
catalog RFGC-numbers of  galaxies from the   $N=490$ sample are
marked by one asterisk (*). The catalog RFGC-numbers of  galaxies
from the $N=441$ sample are marked by an additional asterisk and are
listed in the base sample as (**).

\renewcommand{\baselinestretch}{1.2}
\setcaptionwidth{\linewidth}%
\setcaptionmargin{0mm} %
\onelinecaptionstrue
\captionstyle{nonumber}
\medskip
\begin{longtable}{lllllllllllll}
\caption{}\\
 1**    & 6**    & 16     & 18     & 25     & 31     & 34**   & 46     & 58     & 73**   & 81     & 88     & 96\\
 99**   & 106    & 113**  & 119**  & 122**  & 123    & 124**  & 132    & 136**  & 155    & 161**  & 164**  & 166** \\
 175    & 176**  & 187    & 193**  & 195    & 207**  & 210    & 213**  & 225**  & 229    & 234**  & 237**  & 239**\\
 248    & 255**  & 261**  & 267    & 278**  & 282**  & 283    & 292    & 296**  & 300    & 301    & 302**  & 321\\
 322    & 325    & 330**  & 333    & 337    & 342**  & 344**  & 357**  & 365**  & 368    & 371**  & 374**  & 377 \\
 381*   & 385**  & 389**  & 392    & 393    & 403    & 415    & 430*   & 433    & 438**  & 446**  & 461    & 463**\\
 465**  & 467**  & 483    & 484**  & 485**  & 486    & 493    & 500*   & 504**  & 505**  & 509**  & 510**  & 511** \\
 512    & 513    & 517**  & 523    & 527    & 529    & 531**  & 537    & 539    & 542    & 544**  & 557    & 560**\\
 569    & 598**  & 603**  & 604**  & 615    & 620**  & 622**  & 625**  & 626**  & 627**  & 631    & 634**  & 642\\
 653**  & 660    & 666    & 674**  & 676**  & 679    & 687**  & 693**  & 695**  & 697**  & 701    & 705    & 719** \\
 720    & 722**  & 730    & 744**  & 746**  & 754    & 756**  & 766**  & 768    & 769**  & 772    & 778**  & 793\\
 798**  & 809*   & 813    & 826**  & 827    & 828*   & 831    & 835    & 844*   & 849**  & 854    & 855**  & 858**\\
 863    & 869*   & 871    & 877**  & 880    & 888**  & 900**  & 903**  & 911**  & 912*   & 925**  & 928    & 934 \\
 940    & 942    & 944**  & 946    & 950    & 972    & 975    & 979    & 988**  & 990    & 995    & 1000** & 1005\\
 1010   & 1013   & 1015*  & 1016   & 1021   & 1029*  & 1045   & 1046   & 1050   & 1051** & 1055** & 1059   & 1065\\
 1083   & 1091** & 1094   & 1095** & 1099** & 1107   & 1109** & 1112** & 1113*  & 1114** & 1124   & 1129   & 1132**\\
 1133** & 1135** & 1140** & 1143** & 1147** & 1148   & 1149** & 1150   & 1155   & 1157   & 1162   & 1169   & 1170\\
 1171   & 1172** & 1184** & 1189*  & 1195** & 1196** & 1200** & 1201   & 1211** & 1231** & 1234   & 1236** & 1248** \\
 1256   & 1259** & 1275   & 1277** & 1278   & 1284   & 1285*  & 1291** & 1293** & 1298** & 1300** & 1305   & 1313*\\
 1322** & 1325** & 1329** & 1333   & 1339** & 1344   & 1346*  & 1348** & 1355   & 1357   & 1358   & 1359** & 1362*\\
 1363   & 1365** & 1366** & 1374   & 1375** & 1383** & 1384*  & 1385   & 1387** & 1392   & 1394** & 1399** & 1406\\
 1407   & 1412   & 1413   & 1417** & 1420** & 1424   & 1425** & 1433   & 1434** & 1435** & 1439** & 1440** & 1443**\\
 1446** & 1461*  & 1462** & 1468** & 1470** & 1486** & 1490** & 1502** & 1504** & 1514** & 1522** & 1527** & 1530**\\
 1549   & 1551   & 1552   & 1553** & 1555** & 1556** & 1560** & 1561** & 1562   & 1563   & 1567** & 1568** & 1569**\\
 1572   & 1587** & 1595** & 1596   & 1597   & 1607   & 1620** & 1625** & 1626** & 1627** & 1630** & 1636   & 1637 \\
 1646** & 1648   & 1650** & 1660** & 1664   & 1670** & 1672** & 1674*  & 1685** & 1687   & 1691** & 1692** & 1696*\\
 1700** & 1710** & 1716** & 1737   & 1739** & 1742   & 1744** & 1745   & 1753** & 1761** & 1766** & 1771   & 1778\\
 1782** & 1783   & 1785   & 1789** & 1791** & 1795** & 1796** & 1807   & 1823   & 1834   & 1837*  & 1843   & 1845\\
 1847   & 1864** & 1871** & 1874** & 1880** & 1886   & 1888** & 1892** & 1918   & 1925** & 1936** & 1940** & 1943\\
 1951   & 1952   & 1953   & 1957*  & 1958** & 1959** & 1973   & 1976** & 1979** & 1992   & 1993** & 1996   & 2000**\\
 2002** & 2004   & 2005   & 2006** & 2010** & 2014*  & 2020** & 2021** & 2026** & 2030   & 2035** & 2037** & 2039\\
 2041   & 2042** & 2045   & 2046   & 2048*  & 2050** & 2051** & 2053   & 2057** & 2058   & 2061** & 2067** & 2069**\\
 2077** & 2078   & 2079** & 2085** & 2106** & 2108   & 2111** & 2118   & 2123   & 2126** & 2127** & 2129** & 2132**\\
 2145** & 2146** & 2148** & 2156** & 2158*  & 2165** & 2173** & 2175   & 2186   & 2187   & 2195*  & 2206** & 2207\\
 2210** & 2211   & 2218** & 2233** & 2243   & 2246** & 2250** & 2253** & 2259** & 2260** & 2266** & 2283*  & 2290**\\
 2295** & 2297** & 2301   & 2308** & 2317** & 2320** & 2322** & 2327** & 2333** & 2339** & 2341   & 2347   & 2350\\
 2357** & 2361** & 2368** & 2378** & 2380** & 2382** & 2387** & 2392   & 2396   & 2398   & 2399** & 2403   & 2415**\\
 2421** & 2428** & 2429** & 2430** & 2443** & 2446** & 2450   & 2453** & 2459** & 2461*  & 2467** & 2468** & 2474**\\
 2482   & 2496** & 2506   & 2510   & 2523** & 2524** & 2526** & 2531** & 2546** & 2548*  & 2549** & 2550   & 2551**\\
 2553   & 2555** & 2556   & 2569** & 2580** & 2581** & 2585   & 2590** & 2592*  & 2597   & 2605   & 2611** & 2617**\\
 2623** & 2631** & 2635** & 2637   & 2642   & 2652** & 2668** & 2676   & 2687** & 2694** & 2699** & 2702** & 2705**\\
 2714   & 2716   & 2725** & 2727   & 2736*  & 2740   & 2750*  & 2751   & 2752*  & 2755*  & 2757** & 2768** & 2774**\\
 2779*  & 2785*  & 2791   & 2795** & 2802** & 2803** & 2805** & 2808*  & 2811** & 2819** & 2821   & 2834   & 2842\\
 2843   & 2853** & 2855   & 2864** & 2872** & 2875** & 2876   & 2881   & 2885** & 2886** & 2897** & 2898   & 2901**\\
 2905   & 2906** & 2908** & 2918** & 2919   & 2923** & 2927** & 2928** & 2929   & 2931** & 2932   & 2953   & 2965 \\
 2966** & 2968   & 2969** & 2979   & 2988** & 2999*  & 3001   & 3007** & 3008   & 3013   & 3027** & 3032   & 3034\\
 3036** & 3037** & 3044** & 3046** & 3054** & 3064   & 3069   & 3076   & 3077   & 3078   & 3079   & 3080   & 3081**\\
 3082   & 3087** & 3094** & 3095** & 3097** & 3103   & 3118** & 3119   & 3125   & 3126   & 3129   & 3132** & 3140\\
 3153*  & 3158   & 3160** & 3164** & 3170   & 3172   & 3184   & 3185** & 3186** & 3190** & 3198   & 3200   & 3208\\
 3212** & 3218** & 3219** & 3232** & 3243** & 3245   & 3256   & 3258** & 3262** & 3274** & 3277** & 3285** & 3289**\\
 3296   & 3297** & 3304** & 3316** & 3322   & 3326   & 3331** & 3337** & 3350   & 3354   & 3356** & 3359** & 3364*\\
 3365** & 3367** & 3371   & 3372   & 3377** & 3378** & 3383   & 3385** & 3397   & 3400   & 3405** & 3414** & 3424\\
 3431   & 3437** & 3438   & 3439   & 3444** & 3452   & 3465** & 3468   & 3481** & 3488   & 3489** & 3491** & 3515**\\
 3516   & 3517   & 3518   & 3519** & 3520** & 3521   & 3522   & 3526   & 3527** & 3532** & 3534   & 3540   & 3555 \\
 3558   & 3561** & 3563   & 3575** & 3580** & 3582   & 3587   & 3596   & 3598   & 3600** & 3601   & 3603   & 3608**\\
 3611   & 3619   & 3622*  & 3628   & 3631** & 3636   & 3643** & 3644** & 3645** & 3651** & 3659*  & 3672   & 3684*\\
 3686   & 3689   & 3697   & 3699** & 3703   & 3707   & 3712** & 3715   & 3723** & 3727** & 3729   & 3740   & 3750\\
 3752*  & 3753** & 3761   & 3769   & 3779   & 3783   & 3792** & 3798   & 3803** & 3804   & 3820** & 3822   & 3824**\\
 3827** & 3828** & 3833** & 3846** & 3854** & 3856   & 3858** & 3879** & 3880** & 3896** & 3901   & 3915** & 3919\\
 3924   & 3928*  & 3930   & 3935** & 3952** & 3953** & 3968   & 3975   & 3977** & 3984*  & 3986** & 3998** & 4003\\
 4010   & 4013** & 4025** & 4028** & 4032** & 4039** & 4046   & 4053*  & 4054   & 4057** & 4066** & 4072** & 4073\\
 4074** & 4076** & 4078** & 4080** & 4081** & 4091** & 4101   & 4106** & 4113** & 4119** & 4123*  & 4131   & 4134**\\
 4137   & 4149** & 4151   & 4163** & 4172   & 4190** & 4199   & 4203** & 4209** & 4214*  & 4230   &        &\\
\end{longtable}
\twocolumngrid

\renewcommand{\baselinestretch}{1.0}

{}

\end{document}